# An Effective Entropy-assisted Mind-wandering Detection System with EEG Signals based on MM-SART Database

Yi-Ta Chen, Student Member, IEEE, Hsing-Hao Lee, Ching-Yen Shih, Zih-Ling Chen, Win-Ken Beh, Student Member, IEEE, Su-Ling Yeh, and An-Yeu Wu, Fellow, IEEE

**Abstract**— Mind-wandering (MW), which usually defined as a lapse of attention, occurs between 20%-40% of the time, has negative effects on our daily life. Therefore, detecting when MW occurs can prevent us from those negative outcomes resulting from MW, such as failing to keep track of course during learning. In this work, we first collect a multi-modal Sustained Attention to Response Task (MM-SART) database for detecting MW. Eighty-two participants' data are collected in our experiments. For each participant, we collect measures of 32-channels electroencephalogram (EEG) signals, photoplethysmography (PPG) signals, galvanic skin response (GSR) signals, eye tracker signals, and several questionnaires for detailed analyses. Then, we propose an effective MW detection system based on the collected EEG signals. To explore the non-linear characteristics of EEG signals, we utilize the entropy-based features in time, frequency, and wavelet domains. The experimental results show that we can reach 0.712 AUC score by using the random forest (RF) classifier with the leave-one-subject-out cross-validation. Moreover, to lower the overall computational complexity of the MW detection system, we apply techniques of channel selection and feature selection. By using the only two most significant EEG channels, we can reduce the training time of the classifier by 44.16%. By performing correlation importance feature elimination (CIFE) on the feature set, we can further improve the AUC score to 0.725 but with only 14.6% of the selection time compared with the recursive feature elimination (RFE) method. By proposing the MW detection engine, current work can be applied to educational scenarios, especially in the era of remote learning nowadays.

——————————— ◆ ———————————

## 1 INTRODUCTION

MIND-WANDERING (MW) is the experience of thoughts not remaining on a single topic, particularly when people are engaged in an attention-demanding task [1]. It has been estimated that the occurrence of MW is between 20%-40% in our daily life [2]. If MW occurs during driving, it might put people in danger. Also, MW will degrade the efficiency of learning if it appears when people are studying [1], especially during the COVID-19 pandemic where on-line learning becomes a necessity. Therefore, an efficient mechanism to analyze and detect MW is of great interest in recent days.

There were several experiments designed to induce MW. In [4], they used the Sustained Attention to Response Task (SART) to analyze the everyday attentional failures and

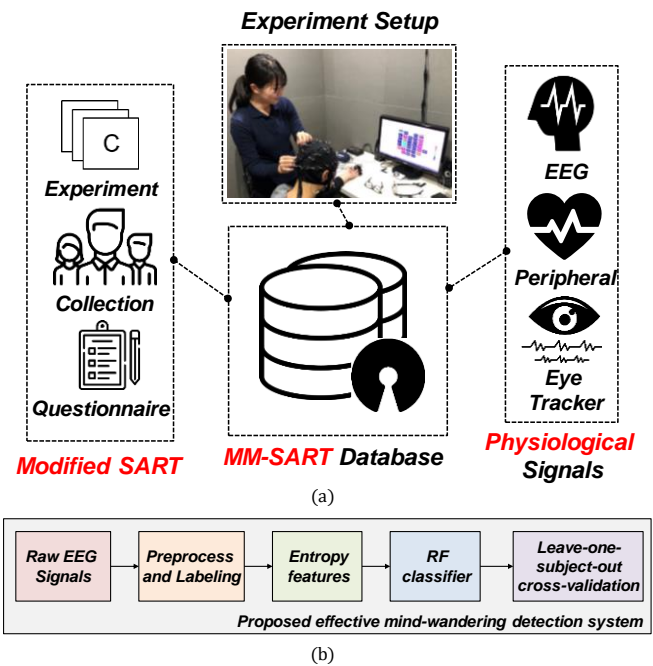

Fig. 1. Overview of (a) proposed MM-SART database, and (b) proposed effective mind-wandering detection system based on EEG signals.

action slips between brain-injured patients and normal participants. The continuous temporal expectancy task (CTET) is introduced to analyze EEG signals between hit and miss during the experiment [5]. In [6], the authors designed the

————————————————

- Y.-T. Chen and W.-K. Beh are with the Graduate Institute of Electronics Engineering, National Taiwan University, Taiwan. E-mail: {edan,kane}@ access.ee.ntu.edu.tw.
- H.-H. Lee is with the Department of Psychology, National Taiwan University, Taiwan. E-mail: hsinghaolee@gmail.com.
- C.-Y. Shih is with the Computer Science, University of Michigan, Ann Arbor, Michigan, United States. E-mail: cys@access.ee.ntu.edu.tw.
- Z.-L. Chen is with the Graduate Institute of Brain and Mind Sciences, National Taiwan University College of Medicine, Taiwan. E-mail: annie0191@hotmail.com.
- S.-L. Yeh is with the Department of Psychology, National Taiwan University, Taiwan. E-mail: suling@g.ntu.edu.tw.
- A.-Y. Wu is with the Graduate Institute of Electronics Engineering, National Taiwan University, Taiwan. E-mail: andywu@ntu.edu.tw.

experiment by collecting eye gaze data and galvanic skin response of an Affectiva Q sensor while participants were reading texts. The authors of [7] designed a simulated driving experiment to detect MW by EEG signals and labeled the data by auditory probes. The aforementioned studies have shown that MW can be induced by proper experimental designs, and states of attention can be captured by certain physiological measurements.

Since MW is a mental state defined as the lapse of attention, it is highly correlated with the internal activities of our brain. As a result, researchers often study the relationship between electroencephalogram (EEG) signals and MW. For example, event-related potentials (ERP) were used to explore the effect of MW on processing relevant or irrelevant events [3]. In [4], the authors observed increased power in the α band during MW periods. In [5], the non-linear regression model was used to predict MW based on the mean power value of each electrode. In [6], the concentrated state was determined by using the power ratio between β and θ bands as a parameter. Most of the studies used band power and ERPs as the major features. However, note that EEG is a multidimensional, non-stationary, and non-linear time series. Therefore, by using only statistical features in time, frequency, and wavelet domains, which are linear features, one cannot be able to capture all EEG characteristics, especially for those non-linear features. Hence, it is desirable to explore new non-linear methods to extract useful information from EEG data.

In recent years, several non-linear features of EEG were proposed. For example, the authors employed permutation entropy to perform complexity analysis of Alzheimer's disease [7]. In [8], Wavelet entropy was applied in discriminating patients with attention deficit hyperactivity disorder (ADHD) from healthy controls, and a 96% accuracy was achieved in this task. In [9], Fractal Dimension, as a type of complexity feature, was also treated as a biomarker for dementia. These related works have demonstrated that the complexity information of EEG can perform well in the classification for those patients under study. However, none of the work has adopt entropy in characterizing human attentional functions. Hence, it is desirable to extract entropy-based features to detect MW using EEG data.

Besides, due to the multi-channel characteristics and many extracted features of EEG, the overall computational complexity of the system becomes extremely high. Therefore, several simplification methods were proposed to lower the total system complexity. In [10], the authors used features with $p$-values smaller than 0.05 in the emotion recognition task to reduce the total number of features. This can not only reduce the computational complexity of the classifier, but also help analyze the dominant features to support their study. The authors of [11] proposed multiscale permutation entropy (MPE) to reduce the complexity of original multiscale entropy (MSE) in [12]. Those simplified techniques can lower the total complexity and further improve the overall efficiency of the system.

In order to build a MW detection system, several issues need to be considered:
1. **A comprehensive database for MW detection:** There are some desirable features of a complete MW database: First, the generalization of the database is important. Second, multi-modality is required to evaluate the effectiveness of different physiological signals. Third, large number of subjects is critical to the confidence of detection performance. There are already several experiments designed to collect data for detecting MW [2] [5] [6] [7] [17]. Nevertheless, they cannot cover all three desired features.
2. **Exploration of non-linear EEG features for MW detection:** Many studies have focused on the processing of EEG in MW detection. However, most of them applied basic EEG features on the time or frequency domain to reveal the differences between MW and non-MW. Due to the non-linearity of EEG, the effectiveness of non-linear features, such as entropy-based features, should be explored for detecting MW.
3. **Efficient computations of EEG-based MW detection system:** While dealing with EEG signals, the total number of features is relatively large due to the many channels of EEG. Moreover, the computational complexity of several entropy features is $O(n^2)$. Hence, if we directly extract and apply all EEG features to the MW detection system, the overall computational complexity will be a big burden. Hence, a lighter engine reliving the computational complexity should be proposed.

To tackle the above issues, we collect a new multi-modal Sustained Attention to Response Task (MM-SART) database, and design an effective infrastructure of detecting MW based on EEG signals. The main contributions of this paper are as follows:
1. **Multi-modal Sustained Attention to Response Task (MM-SART) database:** To support our research on MW detection based on physiological signals, we design a new Multi-Modal Sustained Attention to Response Task (MM-SART) database. Multi-modality physiological signals are collected in a controlled environment to exclude external factors. Moreover, the total subject number is critical to train the MW detection system. Up to now, we have collected data from 82 participants to support our analysis. To the best of our knowledge, the MM-SART database is the first database with most modalities and the largest number of subjects for MW detection. Moreover, the MM-SART database will be open-sourced. For researchers who are interested in doing experiments for MW detection, they can easily access the MM-SART database (URL: http://mmsart.ee.ntu.edu.tw/).
2. **Application of entropy-based features to MW detection:** We propose to extract effective non-linear information of EEG based on entropy-based features. We analyze entropy-based features on time, frequency, and wavelet domain, respectively. From the experimental results, we show that the extracted entropy-based features are complementary to traditional linear features (e.g., statistical and band power features). By utilizing the new entropy-domain features, we can reach a 0.670 F1 score, 0.318 Kappa score, and 0.712 Area-under-Curve (AUC) score in the leave-one-subject-out cross-validation Comparing with the performance of utilizing only the basic statistical features, which is 0.630 F1 score, 0.237 Kappa score, and 0.677 AUC, we can improve our performance by 0.04 F1 score, 0.081 Kappa score, and 0.035 AUC.

3. **Improvement of computational efficiency on the EEG-based MW detection system:** To lower the overall computational complexity of training the MW detection system, we apply methods of channel selection and feature selection. We firstly apply the AUC-based channel selection method to reach the optimal point between the number of channels and the AUC score. Secondly, we propose a correlation importance feature elimination (CIFE) method based on the Random Forest (RF) classifier to select the most significant features. By applying the only two most significant EEG channels to the MW detection system, we can reduce the training time of the classifier by 44.16% but with only 0.016 degradation of AUC score. By performing CIFE on the feature set, we can further improve the AUC score to 0.725 but with only 14.6% of the selection time compared with the recursive feature elimination (RFE) method.

The rest of the paper is organized as follows. In Section II, the MM-SART database is introduced. The proposed effective MW detection system based on EEG is presented in Section III. In Section IV, the experiment result of the effective MW detection system is shown. Channel selection and feature selection of the EEG-based MW detection system are introduced in Section V. Finally, the conclusions are drawn in Section VI.

## 2 COLLECTION OF MIND-WANDERING DATASET AND TASK DESCRIPTION

### 2.1 Background

To detect and analyze the process of MW, a proper experiment with higlidity should be proposed. Moreover, the source used to detect MW is important. MW is related to several neural processes [13], such as increased activities in the default mode network (DMN), suppressed activities within the anti-correlated (task-positive) network (ACN), as well as other changes in neuromodulation. Previous studies have utilized these brain connections to build up a model to detect states of attention [13] and consciousness [14]. However, using functional magnetic resonance imaging (fMRI) is less flexible and portable, which cannot be easily utilized in daily situations. Therefore, detecting MW by multi-modality physiological signals is more applicable to educational scenarios as the current study has done.

In our experiments, we adopt a modified version of SART. We use a pseudo-random probe-based method to access the mental state of subjects. Moreover, we collect a multi-modality database to support various kinds of research. The details of the database are in the appendix. The website of the complete MM-SART database can be found in: http://mmsart.ee.ntu.edu.tw/.

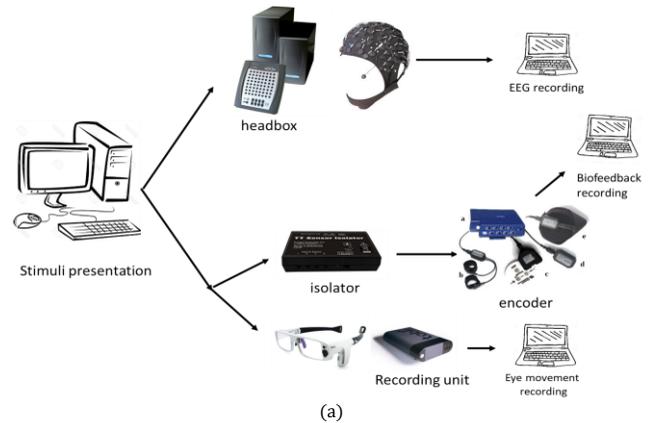

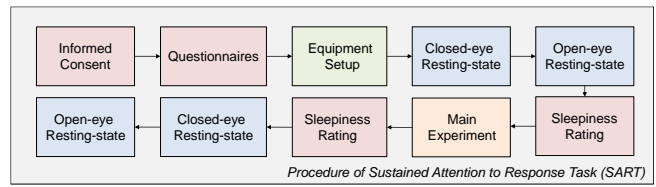

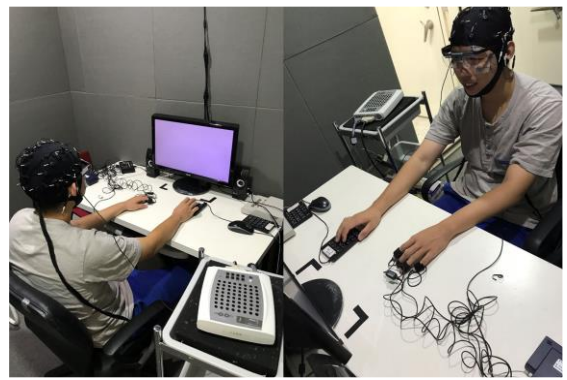

Fig. 2. (a) Data collection and synchronization among Neuroscan, Tobii Eye Glasses and ProComp Infiniti bio sensor system. (b) Procedure of the SART experiment. (c) The demonstration of a participant performing the task.

### 2.3 Participants

Eighty-two participants are recruited in the current study. Five participants are excluded from the data analysis due to technical issues. Therefore, we end up with 77 participants (age range: 20-33 years old, 40 females). All of the participants are right-handed and free from psychological and neurological disorders. They all have normal or corrected-to-normal vision. The experiment is approved by the Research Ethics Committee at National Taiwan University (NTU REC: 201812HM004) and executed with the appliance to the guidelines.

### 2.4 Apparatus

All stimuli are presented in a gray background on an ASUS 22'' LED monitor with a spatial resolution of 1920 × 1080 pixels. EEG, eye tracker signals, and physiological signals are recorded by 3 systems, respectively. Stimuli are presented with E-prime (Psychology Software Tools, Pittsburgh, PA, USA), and triggers are also sent by E-prime and synchronized with a DB-25 connector.

EEG data is recorded with Neuroscan (El Paso, TX, USA) with 32-channel Quick-cap (AgCl electrodes). The recordings are originally referenced to the left mastoid (M1), and are re-referenced to the average of the left and right mastoid (M2) offline. Vertical electrooculogram (V-EOG) is recorded from participants' left eye with two electrodes (one placing on approximately 2cm above the left eye, and the other was 2cm below the left eye). Horizontal electrooculogram (H-EOG) is recorded with pairs of electrodes placing at 2cm away from the left and right eye respectively. Before starting the experiment, the impedances of all electrodes are kept under 5k$\Omega$ to ensure the quality of data. EEG and EOG signals are amplified by the SynAmps using a 0.05–100Hz bandpass and continuously sampled at 1000 Hz per channel for offline analysis.

All participants' heart rate, skin conductance, skin temperature, and respiration data are recorded by the Pro-Comp Infiniti (ProComp Infiniti of Thought Technology Ltd) at 2048 Hz and downsampled to 256 Hz while exporting the data. In addition, their eye movements data are recorded by Tobii Eye Glasses 2 (Tobii Technology, Danderyd, Sweden) with the sampling rate of 100 Hz.

### 2.5 Stimuli and Design

Participants are seated in a sound-attenuated room with their eyes approximately 80cm from the monitor. They are instructed to do the SART proposed in [15], and [16].

In this task, each block includes 25 trials and a probe at the end of the block. Participants are instructed to press number 9 on the number-pad with their right hand to initiate a block. Each block is embedded with 25 English letters (A-Y) in a pseudo-random order with one target letter (i.e., letter C, and the target probability was 4%), which appears pseudo-randomly at one of the trials between the 6th and 15th trial in a block.

Participants are instructed to press number 8 with their right hand as soon as possible when they catch sight of a non-target letter but to withhold their response when they see the target letter "C". Each letter is presented for 2000ms or until the participant responds. The inter-trial interval (ITI) varies with the reaction time (RT) of participants so that each trial (including ITI) lasts for 2000ms. For example, if the participant's response time is 300ms, then the ITI would be 1700ms to equate the duration of each trial. There are 40 blocks in total.

### 2.6 Procedure

The procedure of the overall experiment is shown in Fig. 2(b). After signing the informed consent, participants are instructed to fill in the questionnaires. After that, they are equipped with bio-sensors (on their left-hand fingers), EEG cap, and Tobii Eye Glasses 2. Next, participants are recorded during a 3-minute closed-eye and open-eye resting-state. Then, participants are instructed to practice for 3 blocks to make sure that they understand the SART. Before the formal experiment starts, participants are asked to rate their sleepiness on a 4-point Likert scale (from 1: very alerted to 4: very sleepy) of their current state. Participants are told to do the task at their own pace and are allowed to rest at the end of each block. After the formal experiment ends, participants are asked to rate their state of sleepiness again on the 4-point Likert scale, followed by closed-eye and open-eye resting-state signals recording.

### 2.7 Self-assessment of Participants

At the end of the block, a probe pops out and asks participants to classify the content of their thoughts with the question "What was in your mind just now?" with five options (1. Focusing on the task; 2. Thinking of the task performance; 3. Distracted by task-unrelated stimuli; 4. Thinking of things unrelated to the task; and 5. Nothing in particular). After classifying their thoughts, another rating question asks participants to subjectively rate their state of focus from 1 (completely wandering) to 7 (very focused) at the moment before seeing the probe. Participants are told that they should respond honestly and that there is no correct answer for the probe and the rating questions.

## 3 PROPOSED EEG-BASED MIND-WANDERING DETECTION SYSTEM DESIGN

In this work, we focus on the designing of a MW detection system based on the EEG signal. The use of multi-modality signals can be extended based on this initial study of the MM-SART database. From several related works, the processing flow of EEG can be separated into few steps as shown in Fig. 3. The details of each building block will be described as follows.

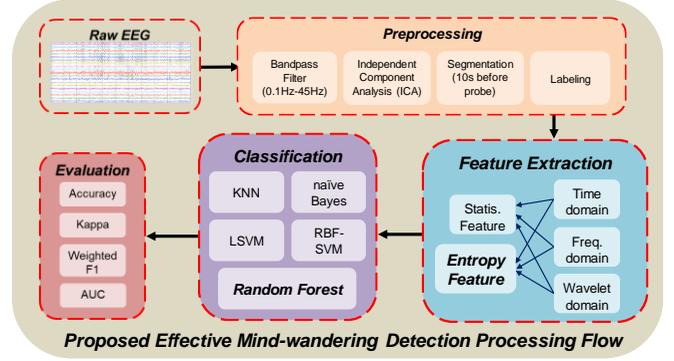

Fig. 3. The processing flow of EEG signals for the MM-SART database.

TABLE 1
BASIC FEATURES AND ENTROPY-BASED FEATURES IN THREE DIFFERENT DOMAINS

| Feature sets | Time | Frequency | Wavelet |
|---|---|---|---|
| Statistical | Mean, Mean Power, $1^{st}$ diff, $2^{nd}$ diff, Hjorth complexity | Power spectral density of θ, α, β, γ bands | Mean power, Mean, Standard deviation, a ratio of absolute mean values of adjacent bands Mean power, Mean, Standard deviation, a ratio of absolute mean values of adjacent bands |
| Entropy | MSE, MPE, MDE, MFDE | Spectral entropy | MPE, MDE, MFDE |

## 3.1 Preprocessing and Labeling

In the preprocessing step, we re-reference all EEG channels according to the average of the M1 and M2 channels. We use a bandpass filter from 0.1 Hz to 45.0 Hz to remove the baseline and high-frequency noise. Eye artifacts are noise when classifying EEG signals [7], and hence independent component analysis (ICA), such as FastICA algorithm [23], is usually applied to the EEG sequences to remove eye artifacts. However, according to [6], eye blink information is useful for detecting MW. Therefore, we propose to process the EEG signals without ICA to utilize the information of eye movements in EEG signals and compare the classification performance between with and without ICA.

Finally, we segment 10-s EEG signals before probes into epochs for each subject to predict the mental state. The labeling of each segmentation is based on self-assessment. In this paper, we consider the 7-point Likert scale as our labeling target to evaluate the subjective MW state. Moreover, trials scoring 4 are removed because they cannot be categorized into either of the two states. After labeling each trial, we remove 21 subjects who always labeled themselves as MW or not MW. By doing so, we can prevent the undefined value of the F1 score with the leave-one-subject-out cross-validation.

## 3.2 Feature Extraction

We extract basic statistical and entropy-based features, which are summarized in Table 1. As for the entropy-based features, we extract the multiscale entropy (MSE) [16], the multiscale permutation entropy (MPE) [15], the multiscale dispersion entropy (MDE) [24], and the multiscale fluctuation-based dispersion entropy (MFDE) [25] to capture the complexity of EEG signals in different scales on time and wavelet domains. These non-linear entropy-based features are introduced as follows. Note that the "multiscale" means coarse-graining process before entropy calculation.

### 3.2.1 Multiscale Entropy (MSE) [12]

The extraction of multiscale entropy (MSE) consists of two steps. The first step is the coarse-graining process. The coarse-graining process will average the data points within a non-overlapping window of length τ, where τ is called the scale factor. Each element of the coarse-grained series, $y_j^\tau$, is represented as:

$$y_j^\tau = \frac{1}{\tau}\sum_{i=(j-1)\tau+1}^{j\tau} x_i. \quad (1)$$

The second step of MSE extraction is the sample entropy calculation on the coarse-grained time series. This step aims to capture the probability of new pattern generated in the coarse-grained time series. The higher value the sample entropy is, the higher probability of new pattern generation. The sample entropy calculation step is defined as follows:

$$SampEn(y^\tau, m, \gamma) = -\ln \frac{n^{m+1}}{n^m}, \quad (2)$$

where $y^\tau$ represents the coarse-grained time series, $n^m$ is the number of matched patterns with dimension $m$ in $y^\tau$, and $\gamma$ is the maximum matching tolerance.

### 3.2.2 Permutation Entropy (PE) [15]

Permutation entropy (PE) is based on the counting of ordinal patterns that describe the up-and-down in the signals. The permutation pattern is denoted as a motif that uses relative order to indicate different kinds of amplitude variation of the signals. Specifically, with the pattern of dimension $m$, there are $m!$ distinct permutation patterns $\{\pi_1, \pi_2, \dots, \pi_{m!}\}$ in the signal $x$ of length $N$. The probability of each pattern is defined as:

$$p(\pi_j) = \frac{\#\{i|0<i \leq N-m, (x_{i+1},\dots,x_{i+m}) \text{ has type } \pi_j\}}{N-m+1}, \quad (3)$$

and the permutation entropy value is calculated based on the Shannon's definition of entropy as:

$$PermEn(x, m) = -\sum_{j=1}^{m!} p(\pi_j) \ln p(\pi_j). \quad (4)$$

### 3.2.3 Dispersion Entropy (DE) [17]

Dispersion entropy (DE) can detect noise bandwidth, simultaneous frequency, and amplitude change. The dispersion entropy calculation consists of four steps. In the first step, the time series $x = \{x_1, x_2, \dots, x_N\}$ is mapped to $c$ classes. By employing the normal cumulative distribute function (NCDF)

$$y_j = \frac{1}{\sigma\sqrt{2\pi}}\int_{-\infty}^{x_j} e^{\frac{-(t-\mu)^2}{2\sigma^2}} dt, \quad (5)$$

we can map $x$ into $y = \{y_1, y_2, \dots, y_N\}$. Next, we use a linear algorithm to map each $y_j$ to $z_j^c$ according to

$$z_j^c = round(c \cdot y_j + 0.5), \quad (6)$$

where $z_j^c$ is an integer from 1 to $c$.

In the second step, the embedding vector $z_i^{m,c}$ with embedding dimension $m$ and time delay $d$ is created according to

$$\mathbf{z}_i^{m,c} = \{z_i^c, z_{i+d}^c, \dots, z_{i+(m-1)d}^c\}, i = 1,2,\dots,N-(m-1)d. \quad (7)$$

The time series $\mathbf{z}_i^{m,c}$ is mapped to the dispersion pattern $\pi_{v_0 v_1 \dots v_{m-1}}$, where $v_0 = z_i^c, v_1 = z_{i+d}^c, \dots, v_{m-1} = z_{i+(m-1)d}^c$. The number of possible dispersion patterns is equal to $c^m$.

In the third step, the probability of each pattern is calculated as:

$$p(\pi_{v_0 v_1 \dots v_{m-1}}) = \frac{\#\{i|i < N-(m-1)d, \mathbf{z}_i^{m,c} \text{ has type } \pi_{v_0 v_1 \dots v_{m-1}}\}}{N-(m-1)d}. \quad (8)$$

In the last step, the dispersion entropy with embedding dimension $m$, time delay $d$, and the number of classes $c$ is defined as follows:

$$DispEn(x, m, c, d) = -\sum_{\pi=1}^{c^m} p(\pi_{v_0 v_1 \dots v_{m-1}}) \ln p(\pi_{v_0 v_1 \dots v_{m-1}}). \quad (9)$$

### 3.2.4 Fluctuation-based Dispersion Entropy (FDE) [18]

Fluctuation-based dispersion entropy (FDE) is more stable than dispersion entropy over the irrelevant local trend. The main difference between FDE and DE is the second step mentioned in the dispersion entropy. The FDE considers the differences between adjacent elements of dispersion patterns. Thus, each element in fluctuation-based dispersion patterns changes from $-c+1$ to $c-1$, and there are $(2c-1)^{m-1}$ possible fluctuation-based dispersion patterns. The other calculation step of the FDE is the same as that of DE.

## 3.3 Classifier

In related works [19] [20] [21], the authors applied several machine learning algorithms to detect whether or not the subject is MW, such as naïve Bayes (NB), linear support vector machine (SVM), RBF-kernel support vector machine (RBF-SVM), k-nearest neighbors (KNN), and random forest (RF). In this paper, we evaluate and compare the performance among the 5 most common classifiers mentioned above for the proposed EEG-based MW detection system.

## 3.4 Evaluation Metric

The generalizability of the data is critical in scientific research. If the classifier has already seen some data from a specific subject, overfitting to the individual may happen and jeopardize the external validity of the model. To consider generalizability, several works have suggested applying leave-one-subject-out cross-validation to evaluate the performance. Additionally, appropriate metrics should be used to compare the performance between the methods. In this paper, we apply three evaluation metrics, which are the weighted F1-score, Cohen's Kappa coefficient, and area under ROC curve to compare the performance between methods.

### 3.4.1 Weighted F1-score (F1)

When the data number within each class is imbalanced, the classifier will tend to predict the label of the majority class, which will gain high accuracy score. Compared to accuracy, F1-score is more reliable by taking the recall and precision into consideration. The F1-score is calculated as follows:

$$F1 = \frac{2 \times Recall \times Precision}{Recall + Precision}, \quad (10)$$

where

$$Recall = \frac{TP}{TP+FN}, Precision = \frac{TP}{TP+FP}, TP = true\ positive\ rate, FP = false\ positive\ rate. \quad (11)$$

Moreover, we need to consider the performance of all the classes. In this paper, we use a modified version of the F1-score, weighted F1-score, to evaluate the system performance. The definition of the weighted F1 score is:

$$F1_{weighted} = P_{MW} \times F1_{MW} + P_{non-MW} \times F1_{non-MW}, \quad (12)$$

where $P_{MW}$ is the number of MW instances and $P_{non-MW}$ is the number of non-MW instances.

### 3.4.2 Cohen's Kappa Coefficient (Kappa, κ) [22]

Cohen's Kappa coefficient stands for the agreement between two raters. It is the proportion of agreement after chance agreement is removed from consideration. The Cohen's Kappa coefficient is calculated as:

$$\kappa = \frac{p_0 - p_e}{1 - p_e}, -1 \leq \kappa \leq 1, \quad (13)$$

where $p_0$ is the relative observed agreement among raters, and $p_e$ is the hypothetical probability of the chance agreement. In our case, we use Kappa to measure the agreement between true labels and predicted labels. The better the detection system performs, the higher the Kappa is.

### 3.4.3 Area Under ROC Curve (AUC)

If the classifier can output the probability of each class, then we can calculate the receiver operating characteristic (ROC) curve. In a ROC curve, the x-axis is the false positive rate, and the y-axis is the true positive rate. By calculating the area under the ROC curve (AUC), we can analyze the effectiveness of the prediction model. The chance level of AUC is 0.5, and an excellent model has an AUC close to 1.

## 4 EXPERIMENTAL RESULT ON EEG-BASED MW DETECTION SYSTEM

To verify and evaluate the EEG data of proposed MM-SART database, several aspects are considered. First, we analyze the overall performance of EEG data between with ICA and without ICA. Moreover, we evaluate the performance among channels to find out the impact of eye movement information. After that, we compare the performance in each category of features and find the most useful features for MW detection. Finally, by utilizing the feature importance metric of RF classifiers, we aim to find the most important features to detect MW.

In this experiment, we only apply 30 EEG channels without the H-EOG and the V-EOG. We first do the aforementioned preprocessing on each channel EEG, such as band-pass filter, re-reference, and labeling. After removing the subjects who have the same label in all trials, 56 subjects are included. We apply random search with 5-fold and 100-hyper-parameter combinations to decide the hyper-parameters of each classifier. The details of the hyper-parameters of each classifier are shown in Table 2.

### 4.1 Performance Comparison Between EEG with and without ICA

In this experiment, we want to observe whether eye movement information in EEG can improve the performance of the MW detection system. We process the EEG signal in two manners, with ICA (w/ ICA) and without ICA (w/o ICA). We then extract a total of 424 features, as shown in Table 1 on each channel. Finally, the extracted features are evaluated by the aforementioned five classifiers.

From Table 3, we can observe that the best performance of both cases is acquired by applying RF classifiers as compared to applying other classifiers. Therefore, in the following experiment, we only focus on the performance of RF classifiers. The best F1-score, Kappa, and AUC of EEG w/ ICA are 0.582, 0.148, and 0.617, respectively. The best F1-score, Kappa, and AUC of EEG w/o ICA are 0.670, 0.318, and 0.712, respectively. Therefore, EEG w/o ICA outperforms EEG w/ ICA by 0.088 for the F1-score, 0.170 for Kappa, and 0.095 for AUC. Therefore, when detecting MW by EEG in the MM-SART database, we can utilize the eye movements information in EEG to improve the performance.

TABLE 2
HYPER-PARAMETERS OF EACH CLASSIFIER

|  | Hyper-parameters |
|---|---|
| NB | - |
| KNN | number of neighbors: 10, weights: 'distance', metric: 'manhattan' |
| Linear SVM | C: 0.01 |
| SVM w/ rbf kernel | gamma: 1e-4, C: 5 |
| RF | number of estimators: 700, max features: 'auto', max depth: 12 |

TABLE 3
PERFORMANCE EVALUATION BETWEEN EEG W/ AND W/O ICA

| Classifier | EEG with ICA | | | EEG without ICA (Compare to w/ ICA) | | |
|---|---|---|---|---|---|---|
|  | F1 | Kappa | AUC | F1 | Kappa | AUC |
| NB | 0.565 | 0.108 | 0.556 | 0.584 (+0.019) | 0.144 (+0.036) | 0.579 (+0.023) |
| KNN | 0.563 | 0.100 | 0.569 | 0.602 (+0.039) | 0.189 (+0.089) | 0.621 (+0.052) |
| L-SVM | 0.543 | 0.059 | 0.546 | 0.618 (+0.075) | 0.213 (+0.154) | 0.648 (+0.102) |
| RBF-SVM | 0.581 | 0.135 | 0.602 | 0.654 (+0.073) | 0.285 (+0.150) | 0.695 (+0.093) |
| RF | 0.582 | 0.148 | 0.617 | 0.670 (+0.088) | 0.318 (+0.170) | 0.712 (+0.095) |

Moreover, we count the number of salient features ($p$-value < .05) among all channels in both cases, as shown in Fig. 4. While comparing w/ and w/o ICA, the number of salient features by w/o ICA data is more than that by w/ ICA in most channels. Therefore, most of the features are more discriminative because of the additional information of eye movements in EEG.

Furthermore, we evaluate each channel AUC by RF classifiers respectively. Each channel is trained and evaluated independently in both cases to observe the impact of eye movement information between channels. As shown in Fig. 5, most of the channels have better performance on EEG w/o ICA except for channel O1. In addition, as shown in Fig. 5(b), channels over the right-frontal area have better performance which is in line with previous clinical findings [23] [24].

In summary, we have shown that the performance of EEG w/o ICA outperforms EEG w/ ICA no matter which classifier is used. Moreover, by analyzing the number of salient features and AUC of each channel, we have shown that MW is is highly related to the right frontal regions.

### 4.2 Performance Analysis on Entropy-based Features

In this section, we analyze the effectiveness on different categories of features. Taking the aforementioned experiment results into consideration, we select RF as our classifier and apply it to EEG-without-ICA data. We first count the number of salient features ($p$-value < .05) among the six categories of features, as shown in Table 1. We then train and evaluate the performance of RF classifier on each category of features, respectively. Furthermore, we concatenate the basic statistical features with three domains of entropy-based features and observe whether the entropy-based features can complement the basic features and improve the performance of the RF classifier.

As shown in Table 4, the entropy-based features in the time domain have the highest percentage of salient features, which is 74.8%. Moreover, the largest number of salient features is also entropy-based features in the time domain, which is 1795, followed by the entropy-based features in the wavelet domain, which is 1260. Therefore, the extracted entropy-based features are more discriminative than the basic features.

The experimental result of training the classifiers by each category of features respectively is shown in Table 5. The best performance among six categories of features happens when using only entropy-based features in the time domain, which has 0.669 F1-score, 0.316 Kappa, and 0.708 AUC. Moreover, the performance reached by only the entropy-based features in the time domain is close to the performance when using all features, which is 0.670 F1-score, 0.318 Kappa, and 0.712 AUC.

When combining statistical features with three domains of entropy-based features respectively, the entropy-based features in the time domain improve performance most, as shown in Table 6. Therefore, by extracting entropy-based features in the time domain, extra information is gained, and the RF classifier can learn better.

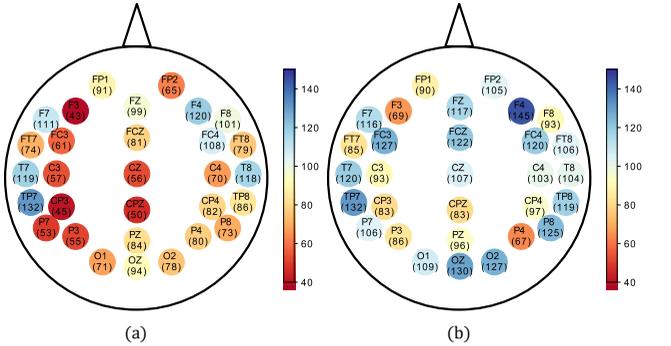

Fig. 4. (a) The number of salient features among all channels on w/ ICA EEG data. (b) The number of salient features among all channels on w/o ICA EEG data.

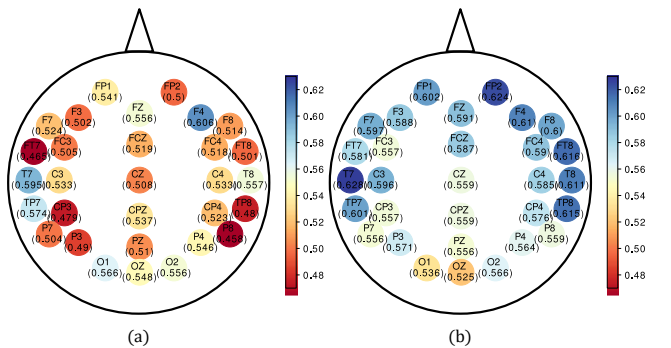

Fig. 5. (a) AUC score of the RF classifier on w/ ICA data among all channels. (b) AUC score of the RF classifier on w/o ICA data among all channels.

TABLE 4
SALIENT FEATURE NUMBER ON SIX CATEGORIES OF FEATURES

| # of salient features/ # of total features (Percentage of salient features) | Time | Frequency | Wavelet |
|---|---|---|---|
| Basic | 107/150 (71.3%) | 85/120 (70.8%) | 354/600 (59%) |
| Entropy | 1795/2400 (74.8%) | 104/150 (69.3%) | 1260/9300 (13.5%) |

TABLE 5
F1/KAPPA/AUC SCORE ON SIX CATEGORIES OF FEATURES BY THE RF CLASSIFIER

| F1/κ /AUC | Time | Frequency | Wavelet |
|---|---|---|---|
| Statistical | 0.642/0.259/ 0.675 | 0.608/0.191/ 0.638 | 0.636/0.249/ 0.684 |
| Entropy | 0.669/0.316/ 0.708 | 0.599/0.173/ 0.642 | 0.495/0.030/ 0.558 |

TABLE 6
F1/KAPPA/AUC ON SIX CATEGORIES OF FEATURES BY THE RF CLASSIFIER

| F1/κ /AUC | Time | Frequency | Wavelet |
|---|---|---|---|
| Statistical | 0.630/0.237/0.677 | | |
| Statistical + Entropy | 0.653/0.283/ 0.702 | 0.640/0.257/ 0.690 | 0.596/0.179/ 0.683 |

In summary, the extracted entropy-based features are indeed discriminative. The performance of the RF classifier can also be improved by applying entropy-based features. We have shown that the entropy-based features are suitable for EEG-based MW detection.

### 4.3 Analysis on the Feature Importance of RF

In this section, we analyze the feature importance derived from the RF classifier. By sorting the feature importance among all features, we can select top-K features as the important feature. By counting the important features of each channel and feature category, we can find the most significant channel or feature category for EEG-based MW detection and strengthen the result of the aforementioned sections.

First, we train the RF classifier by using all-channel EEG data. After training the RF classifier, we extract the feature importance and sort the features with the corresponding importance value. Finally, we select the top-60 features as the analyzed target, which are defined as the significant features. We count and analyze the number of significant features in each channel and category of features.

As shown in Table 7, the channel with the greatest number of significant features is T7. From Fig. 5 (b), the performance of T7 channel is also the best among all channels.

As for the feature category aspect, the experimental result is shown in Table 8. The number of significant features of entropy-based features in the time domain is the greatest and is far more than the other feature categories. This is consistent with the previous result in Section IV.B.

In summary, we analyze each feature from the aspect of importance in RF classifier. The features in channel T7 are more important to the RF classifier among other channels; hence the performance of channel T7 is the best among other channels. Moreover, we have shown that the entropy-based features in the time domain are more important to the RF classifier; hence the performance of entropy-based features in the time domain is also the best among other feature categories.

As a conclusion of this section, we have shown that the effectiveness of overall MW detection by using EEG signals of the MM-SART database. We explore that reserving eye movement information in EEG can improve overall detection performance. Moreover, we extract entropy-based features in the time domain of EEG to complement the traditional statistical features (i.e., mean power, power spectral density), and have shown their effectiveness in detecting MW. By analyzing the importance of each feature in the RF classifier, we further strengthen the importance of channel T7 and entropy-based features in the time domain. Finally, by analyzing the performance of different methods, we can reach 0.670 F1-score, 0.318 kappa score, and 0.712 AUC in the MM-SART database.

## 5 CHANNEL SELECTION AND FEATURE SELECTION ON EEG-BASED MW DETECTION SYSTEM

In this section, we aim to optimize the overall system efficiency by channel selection and feature selection. The former aims to reduce the number of used EEG channels with only a slight degradation of performance. The latter aims to reduce the total number of features in the selected channels to lower the dimension of the classifier's input. Both approaches will help save the computational complexity in training the RF classifier.

TABLE 7
IMPORTANT FEATURE NUMBER OF THE RF CLASSIFER IN EACH CHANNEL

| Channel | Number | Channel | Number | Channel | Number |
|---|---|---|---|---|---|
| Fp1 | 0 | FC4 | 0 | CP4 | 0 |
| Fp2 | 2 | FT8 | 0 | TP8 | 1 |
| F7 | 0 | T7 | 42 | P7 | 0 |
| F3 | 0 | C3 | 0 | P3 | 0 |
| Fz | 0 | Cz | 0 | Pz | 0 |
| F4 | 1 | C4 | 0 | P4 | 0 |
| F8 | 2 | T8 | 1 | P8 | 0 |
| FT7 | 0 | TP7 | 9 | O1 | 0 |
| FC3 | 0 | CP3 | 0 | Oz | 0 |
| FCz | 0 | CPz | 0 | O2 | 0 |

TABLE 8
IMPORTANT FEATURE NUMBER OF THE RF CLASSIFER IN EACH FEATURE CATEGORY

| # | Time | Frequency | Wavelet |
|---|---|---|---|
| Basic | 1 | 2 | 3 |
| Entropy | 49 | 4 | 1 |

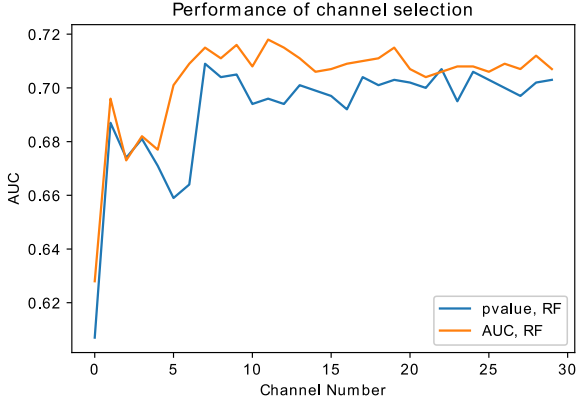

Fig. 6. Channel selection performance evaluation.

TABLE 9
PERFORMANCE (AUC) AND TIME COMPARISON AMONG BOTH CHANNEL SELECTION METHODS

|  | Performance (AUC) | Time (sec) |
|---|---|---|
| Original (30 channels) | 0.712 | 236.23 |
| P-value channel selection (2 channels) | 0.687 (-0.025) | 132.93 (-43.72%) |
| AUC-based channel selection (2 channels) | 0.696 (-0.016) | 131.90 (-44.16%) |

## 5.1 Complexity Analysis of the RF Classifier

The RF classifier [25] is an ensemble model of decision trees. The computational complexity of building a decision tree is $O(Nkd)$, where $N$ is the number of data, $k$ is the number of features, and $d$ is the number of depths of the decision tree. When building RF, two additional parameters need to be decided: the number of trees $m$ and the number of features used to split in each node $k_{sample}$. Therefore, while building decision trees in RF, the computational complexity is reduced to $O(Nk_{sample}d)$. The overall computational complexity of building RF will be $O(mNk_{sample}d)$. Moreover, we have to calculate the additional computational complexity of a random selection of features at each node, which refers to $O(mkd)$. In conclusion, the final computational complexity of building RF classifier is $O(md(Nk_{sample} + k))$. Usually, $Nk_{sample}$ is far larger than $k$, so we can estimate the total computational complexity as $O(mNk_{sample}d)$.

By analyzing the computational complexity of RF, we can conclude that the training time of the RF classifier can be reduced by decreasing $k_{sample}$, $d$, and $m$. However, the parameters $d$ and $m$ are decided by the parameters searching to achieve the best performance on the data. Thus, the only method to reduce the computational complexity is to reduce the $k_{sample}$. In our case, $k_{sample}$ is set to be the square root of $k$. In conclusion, the way to improve the efficiency of the RF classifier is to reduce the feature dimension of the data.

## 5.2 Channel Selection and Feature Selection

As stated above, if the dimension of features can be reduced, the time complexity of training the RF classifier can be reduced, too. The number of features is $n_{channel} \times n_{feature}$, where $n_{channel}$ is the number of channels, and $n_{feature}$ is the number of features per channel. Therefore, we can first reduce the number of channels by finding the critical channels for MW detection. Then, we can perform feature selection on the critical channels to reduce the overall number of features. The following will describe the method of both reducing the number of channels and the number of features.

### 5.2.1 Channel Selection

To reduce the number of channels, we apply two methods to select the optimal channels, which is *p*-value channel selection and AUC-based channel selection. The first method is to calculate the *p*-value of all features of each channel. After calculating the *p*-value of each feature, we count the total number of salient features (*p*-value < .05). We then sort the channel with the number of salient features. This method has the advantage of fast calculation without training extra classifiers. However, the salient features are not equal to significant features, which are decided by the feature importance of the RF classifier.

In contrast, we apply the other method to select the optimal channels. We first train the RF classifier per channel. We evaluate the performance of each trained classifier and sort the channel by its performance, such as AUC. Finally, we can select top-K channels as the candidates to train the final classifier. Although this method requires training additional classifiers, it has the advantage of having better performance. If the specific channel itself can perform well by only its own features, we can infer that by selecting the optimal number of such channels, we can find the best

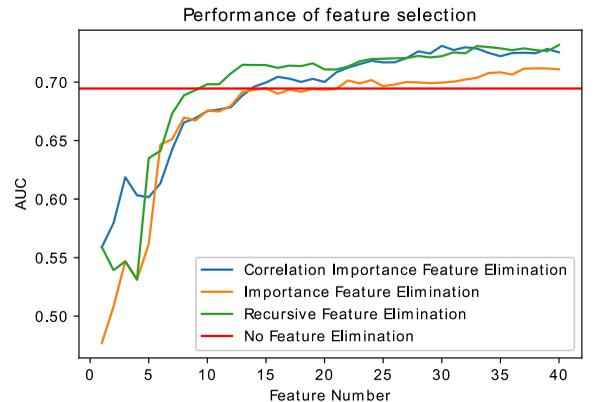

Fig. 7. Feature selection performance evaluation. The red line represents no feature elimination, that is to say, we train the RF classifier with 808 features.

TABLE 10
PERFORMANCE AND TIME COMPARISON AMONG FEATURE SELECTION

| Feature number | | AUC (%) | Overall Time (sec) | Selection Time (sec) |
|---|---|---|---|---|
| 808 | | 69.4 | 177.02 | |
| 40 | RFE | 73.2 (+3.8) | 424.76 | 308.44 |
| | IFE | 71.1 (+1.7) | 157.97 | 41.15 |
| | CIFE | 72.5 (+3.1) | 164.58 | 45.65 |
| 11 | RFE | 69.8 | 441.72 | 321.81 |
| 15 | IFE | 69.5 | 161.46 | 41.79 |
| 14 | CIFE | 69.6 | 172.18 | 47.14 |

TABLE 11
SELECTED FEATURES AFTER DIFFERENT FEATURE SELECTION METHODS

| Method (Number) | Feature Name |
|---|---|
| RFE (11) | T7_PSD_beta, T7_MDE-20, FP2_Mean, FP2_PSD_beta, FP2_PSD_gamma, FP2_cD4-WL-MeanPower, FP2_cD4-WL-RAM, FP2_MFDE-17, FP2_SpecEnt_beta, FP2_SpecEnt_gamma, FP2_cD7-WL-Ent |
| IFE (15) | T7_PSD_beta, T7_MFDE-19, T7_MFDE-20, T7_MDE-18, T7_MDE-20, T7_WL-MFDE-cD7-14, FP2_PSD_beta, FP2_PSD_gamma, FP2_cD4-WL-MeanPower, FP2_cD4-WL-STD, FP2_cD4-WL-RAM, FP2_SpecEnt_beta, FP2_SpecEnt_gamma, FP2_cD7-WL-Ent, FP2_cD7-WL-SpecEnt |
| CIFE (14) | T7_FirstDiff, T7_HjComp, T7_PSD_beta, T7_PSD_gamma, T7_MFDE-1, T7_WL-MFDE-cD7-14, FP2_Mean, FP2_FirstDiff, FP2_PSD_theta, FP2_PSD_beta, FP2_PSD_gamma, FP2_MPE-1, FP2_cD7-WL-Ent, FP2_cD7-WL-SpecEnt |

trade-off between performance and computational complexity.

### 5.2.2 Feature Selection
After reducing the number of channels, we further lower the number of features in each channel to lighten the model complexity. In [30], the feature selection method has been shown to improve learning performance, increase computational efficiency, decrease memory storage, and build better generalization models. Therefore, we propose a specific feature selection method: correlation importance feature elimination (CIFE) for RF classifiers. The CIFE contains two steps: unsupervised correlation clustering and supervised importance rejection.

First, we calculate the correlation between pairs of features with the following equations,

$$\rho(X, Y) = \frac{\sum_i (x_i - \bar{x}_i)(y_i - \bar{y}_i)}{\sqrt{\sum_i (x_i - \bar{x}_i)^2 \sum_i (y_i - \bar{y}_i)^2}}, \quad (14)$$

where X, Y are two different features, and $i$ and $j$ are the $i^{th}$ and $j^{th}$ data respectively. After that, features with correlations higher than $\rho_{thres}$ are clustered. Among each cluster, only one feature will be selected as a representative. Therefore, after correlation clustering, the number of remaining features will equal to the number of clusters. This process can eliminate features that are too similar to highly reduce the number of features.

The second step is to perform supervised importance rejection. Different from unsupervised correlation clustering, supervised importance rejection requires labels to pretrain a RF classifier. After training the RF classifier, top-K remaining features are selected according to the feature importance of the RF classifier. Therefore, the final number of features is K. While comparing with recursive feature elimination (RFE) [26], CIFE is a one-path method which is not necessary to recursively train the classifier. Therefore, the training efficiency of CIFE will be higher than RFE during the selection process.

### 5.3 Experimental Results
To show the improvement of the efficiency in our proposed method, we verify the improvements from the channel selection and feature selection methods, respectively. First, we evaluate the improvement of channel selection and choose an optimal number of channels. Second, we evaluate the improvement of our proposed feature selection method. We implement our experiment on the previously selected optimal channels. Also, we compare different methods by AUC and training time.

The processing of the EEG signals is the same as the previous section. However, considering the issue of efficiency, we eliminate MSE features due to their high time complexity [11].

#### 5.3.1 Experiments on Channel Selection
In order to lower both training and inference complexity of the overall MW detection system, channel selection is necessary. In this experiment, we want to analyze the efficiency of the two channel selection methods, p-value channel selection, and AUC-based channel selection. We first sort the thirty channels according to the number of salient features and AUC scores respectively. We then evaluate the performance of each method by adding one channel at a time. Finally, the AUC metric is used to evaluate the performance.

As shown in Fig. 6, we can see that with only one channel, the performance can only reach 0.61~0.63 AUC. By adding one more channel, the performance of each method reaches close to 0.7 AUC. When comparing two methods, we can see that the performance of AUC-based selection is slightly better than that of p-value selection. As a result, we select two channels to get the best trade-off between computational complexity and performance. The detailed comparison is shown in Table 9. In conclusion, by selecting two critical channels, we lose only 0.016 in AUC but decrease 44.16% of training time in RF classifiers compared to the original 30 channels.

#### 5.3.2 Experiments on Feature Selection
In this section, we aim to further improve the system efficiency with feature selection methods. Therefore, we compare three methods: correlation importance feature elimination (CIFE), importance feature elimination (IFE), and recursive feature elimination (RFE). In the following experiment, we use two critical channels mentioned in the previous section (V.C.1) (i.e., T7 and Fp2). Therefore, the original dimension of the features is 808 (404 (features/channel) × 2 (channels)).

From Fig. 7 and Table 10, when the selected number of features is small, the best feature selection method is RFE, and the worst method is IFE. Moreover, the AUC score of CIFE with 14 features is comparable to the AUC score without feature selection, and the AUC score of RFE with 11 features is comparable to the AUC score without feature selection. However, when the number of features increases, RFE and CIFE converge to almost the same AUC score. In contrast, CIFE performs closely to RFE, but the computational time of CIFE is only 31% of RFE. Therefore, when considering performance and training efficiency, CIFE is the best choice among the three methods.

As shown in Table 11, we list the optimal features selected by these three methods. T7_PSD_beta, FP2_PSD_beta, FP2_PSD_gamma, and FP2_cD7-WL-Ent are chosen by all three methods. Moreover, when observing the category of the selected features, we discover that selected features belong to different categories. Therefore, we conclude that the complementarity between each category of features can help improve the overall performance of the MW detection system based on EEG.

## 6 CONCLUSION

In this paper, we present a multi-modality sustained attention to response task (MM-SART) database. We also propose a framework to detect MW based on the EEG signals collected in the MM-SART database. In our framework, entropy-based features can complement traditional EEG features and therefore improve the performance of the overall system. Moreover, by selecting the best two critical channels, T7 and Fp2, and applying correlation importance feature elimination framework for RF classifiers, we can improve the performance and computational efficiency of the MW detection system based on EEG.

## APPENDIX

## 7 DATA COLLECTION AND TASK DESCRIPTION OF THE MM-SART DATABASE: THE DETAILS

(also available in http://mmsart.ee.ntu.edu.tw/)

### 7.1 Participants

Eighty-two participants are recruited in the current study. Five participants are excluded from the data analysis due to technical issues. Therefore, 77 participants (age range: 20-33 years old, 40 females) remain. All of the participants are right-handed and free from psychological and neurological disorders. They all have normal or corrected-to-normal vision. Participants sign the informed consent before the experiment and is rewarded 400 NTD for their participation. The experiment is approved by the Research Ethics Committee at National Taiwan University (NTU REC: 201812HM004) and executed with appliance to the guidelines.

### 7.2 Apparatus

EEG data is recorded with Neuroscan (El Paso, TX, USA) with 32-channel Quick-cap (AgCl electrodes). The recordings were originally referenced to left mastoid (M1), and were re-referenced to the average of the left mastoid and right mastoid (M2) offline. Vertical electrooculogram (V-EOG) is recorded from participants' left eye with two electrodes (one placed approximately 2cm above the left eye and the other 2cm below the left eye). Horizontal electrooculogram (H-EOG) is recorded with pairs of electrodes placed 2cm away from the left and right eyes respectively. Before starting the experiment, the impedances of all electrodes are kept under 5kΩ to ensure data quality. EEG and EOG signals are amplified by the SynAmps using a 0.05–100 Hz bandpass and continuously sampled at 1000 Hz per channel for offline analysis.

Participants' heart rate, skin conductance, skin temperature, and respiration data are recorded by the ProComp Infiniti (ProComp Infiniti of Thought Technology Ltd) at 2048 Hz and downsampled to 256 Hz while exporting the data. Heart rate sensor is tied to the left index finger, skin conductance is measured from the left middle and ring fingers, and skin temperature is measured from the left little finger. In addition, breath inhale and exhale are measured by the belt tied to the abdomen with adequate tightness. Eye movement data are recorded by Tobii Eye Glasses 2 (Tobii Technology, Danderyd, Sweden) with Tobii SDK program sampled at 100 Hz. Before the recording session started, participants are instructed to do a one-point calibration by holding the calibration card (43 mm in diameter) 1 meter straight away to their eyes and fixating at the bullseye (1.5 mm in diameter) until the calibration was done. After that, no further calibration is required given that Tobii Eye Glasses 2 have high tolerance to motor movements, which allow participants to conduct the experiment without constraining head movements on a chin-rest. Participants are asked to provided their diopters of myopia and are equipped with customized short-sighted eye glasses on the Tobii Eye Glasses 2, to avoid eye movement data recording interference from glasses frames. Furthermore, participants are asked not to wear contact lenses during the experiment to avoid unnecessary eye blinks which might contaminate the data.

### 7.3 Experimental Design

Participants are seated in a sound-attenuated room with their eyes approximately 80cm from the monitor. They are instructed to do the SART experiment that was first proposed in [15], and we adopt the version modified from [16].

In this task, each block includes 25 trials and a probe in the end of the block. Participants are instructed to press number 9 on the number-pad with their right hand to initiate a block. Each block is embedded with 25 English letters (A-Y; extending approximately 1cm horizontally and vertically) in a pseudo-random order with one target letter (i.e., letter C, and the target probability was 4%), which appears pseudo-randomly at one of the trials between the 6th and 15th trial in a block. The experimental design is shown in Fig. 8.

TABLE 12
THE SUMMARY OF DATA COLLECTION AND EQUIPMENT

| Recording system | Data contents |
|---|---|
| Neuroscan | EEG |
| Procomp Infiniti | heart beat, skin conductance, skin temperature, respiration |
| Tobii Eye Glasses 2 | eye movement |
| Self-report Questionnaires | MAAS, PSQI |

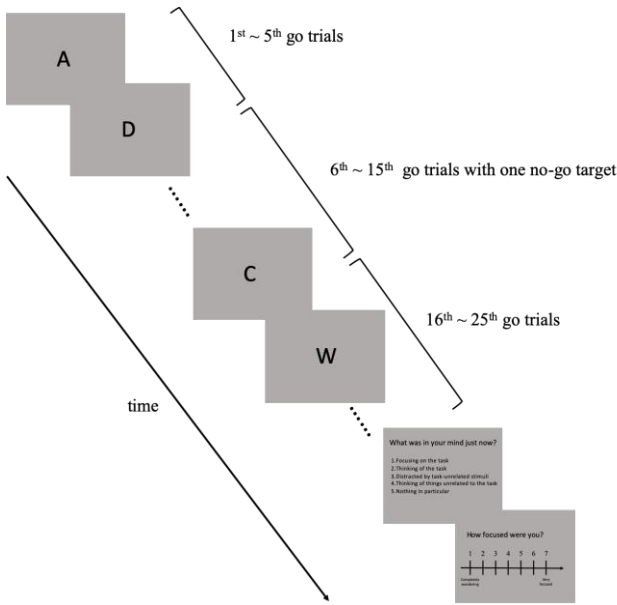

Fig. 8. Experimental design of the SART.

During the task, participants are instructed to press number 8 with their right hand as soon as possible when they catch sight of a non-target letter but to withhold their response when they see the target letter "C". Each letter is presented for 2000 ms or until the participant responds. The inter-trial interval (ITI) varies with the response time of participants so that each trial (including ITI) lasts for 2000 ms. For example, if the participant's response time is 300 ms, then the ITI is 1700 ms to equate the duration of each trial.

At the end of each block, a probe pops out and asks participants to classify "*What was in your mind just now?*" with five options (1. Focusing on the task; 2. Thinking of the task performance; 3. Distracted by task-unrelated stimuli; 4. Thinking of things unrelated to the task; and 5. Nothing in particular). After classifying their thoughts, another probe also asks participants to subjectively rate their state of focus from 1 (completely wandering) to 7 (very focused) at the moment before seeing the previous probe. After that, they could take a break at their own pace or press the number 9 key on the number-pad to proceed to the next block. Participants are told that they should respond honestly and that there is no absolute correct answer for the probe and the rating questions; this is done to avoid possible response bias or social anticipatory effect.

### 7.4 Questionnaires

Two questionnaires are adopted to capture individual differences among participants. The Pittsburgh Sleep Quality Index (PSQI) [27] was adopted to measure participants' sleep quality among the month before the experiment. The PSQI embedded with 19 self-report items with 7 components, including subjective sleep quality, sleep duration, sleep latency, sleep disturbance, sleep efficiency, daytime sleepiness, and medication use. Score in each component ranges from 0 to 3, with 3 indicating the worst sleeping quality. We adopt the PSQI as one of our measurement since previous studies have pointed out that sleepiness could wildly drive the occurrence of mind-wandering [28] [29] as well as influence physiological signals [30]. Hence, by assessing sleep quality of participants via the PSQI, we were able to capture individual differences among the variance of physiological signals as well as the frequency that mind-wandering occurs.

Moreover, the Mindfulness Attention Awareness Scale (MAAS) proposed in [31], a 6-point Likert scale with 15 items, is used to measure participants' trait of mindfulness. MAAS has been widely used to assess people's state of self-awareness, which has been shown to be negatively correlated with the occurrence of MW [32] [33]. That is, the higher the trait of mindfulness, the lower the frequency of mind-wandering. These two questionnaires are adopted to measure traits of participants to assess different aspects in addition to physiological signals that might contribute to detecting mind-wandering.

### 7.5 Procedure

After signing the informed consent, participants are instructed to fill in the questionnaires. After that, their left hand is disinfected with alcohol by the experimenters and equipped with the sensors as well as the EEG cap. After ensuring the impedance of all electrodes were under $5k\Omega$, participants are instructed to do the one-point calibration for Tobii Eye Glasses 2. Next, participants are instructed to close their eyes without thinking about anything to record 3-minute close-eye resting-state signals, followed by a 3-minute open-eye resting-state recording while fixating their eyes at the black fixation cross (extending 1cm vertically and 1.5cm horizontally) on the grey screen. After the resting-state signals recording, participants are told about the task contents and the definitions of the thought contents of the probe. They are instructed to practice for 3 blocks to make sure that they understand the task while the experimenters monitor their performance to make sure that they understand the task rules. Before the formal experiment starts, participants are asked to rate their current state of sleepiness on a 4-point Likert scale (from 1: very alerted to 4: very sleepy).

Participants are told to do the task on their own pace and are allowed to take a break during the end of a block. While participants are performing the task, their performance is monitored by the experimenters outside the laboratory room, where data of each equipment is monitored by an individual computer. After the formal experiment ends, participants are asked to rate their state of sleepiness again on a 4-point Likert scale, followed by the close-eye resting-state signals recording and open-eye resting-state signals recording. The entire experiment takes around 1.5 to 2 hours to finish.

TABLE 13
THE BEHAVIORAL PERFORMANCE IN THE TASK

| Mean (STD) | successful stop | fail-to-stop | rating focused | rating wandering |
|---|---|---|---|---|
| RT (ms) | 412.16 (63.37) | 375.8 (58.46) | 386.07 (78.79) | 395.67 (90.77) |
| RTCV | 0.19 (0.09) | 0.21 (0.1) | 0.21 (0.09) | 0.35 (0.2) |

TABLE 14
SCORES OF THE QUESTIONNAIRES

|  | PSQI | MAAS |
|---|---|---|
| mean | 5.81 | 59.26 |
| sd | 2.98 | 9.64 |

### 7.6 Behavioral Performance

We briefly summarize the behavioral performance here. In the current study, successful stop rate (withhold response while seeing target letter "C") is 73.18%. In the reaction time (RT) and reaction time coefficient of variation (RTCV) analysis, we compare the performance of the 5 trials preceding the target letter "C" and the 5 trials preceding the probe across participants. Precisely, we extract a 10-second time window, which is the same as the classification analysis, to compare the RT and RTCV in objective/subjective focused and wandering states, as shown in TABLE 13. Significant slower RTs are found in successful stop conditions compared to fail-to-stop conditions ($p < .001$). Additionally, a trend for smaller RTCVs in the successful stop condition compared to those of the fail-to-stop condition is found ($p = .065$). On the contrary, there is no significant difference in RT when comparing subjectively rating focused compared to rating wandering ($p = 0.199$). However, a significant larger RTCV is found in the rating wandering condition compared to that of the rating focused condition ($p < 0.001$).

### 7.7 Results of the Questionnaires

Scores of the questionnaires are summarized in Table 14. A significant negative correlation between the MAAS and PSQI was found ($r = -0.56$, $p < .001$). In addition, there is a significant positive correlation between successful stop rate (proportion of successfully withhold response while viewing target letter) and MAAS scores in males ($r = 0.33$, $p = 0.043$), but not in females ($r = -0.01$, $p = 0.96$).

### ACKNOWLEDGMENT

The authors wish to thank the participants that engaged in the data collection of MM-SART. This research was supported in part by the Ministry of Science and Technology of Taiwan (MOST 106-2221-E-002-205-MY3, 109-2622-8-002-012-TA).

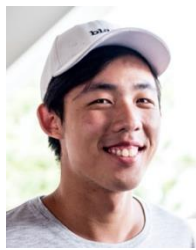

**Yi-Ta Chen (S'17)** received the B.S. degree in electrical engineering from the National Taiwan University, Taipei, Taiwan, in 2017. He is currently working toward the Ph.D. degree in the Graduate Institute of Electronics Engineering, National Taiwan University. His research interests include the machine learning engine for affective computing, biosignal processing and feature extraction for affective computing, and SW/HW co-design for SDN data plane.

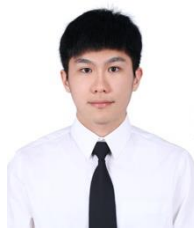

**Hsing-Hao Lee** received his M.S. degree in psychology from National Taiwan University in 2019. His research interests include human attention, cognitive control, consciousness, and aging. He investigates these topics with various methodologies, including fMRI, EEG, eye-tracker, bio-physiological signals as well as behavioral experiments.

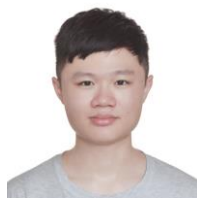

**Ching-Yen Shih** received the B.S. degree in Electrical Engineering from National Taiwan University in 2018. He is currently majoring in Computer Science at University of Michigan. His research interests include computer vision, natural language processing, affective computing.

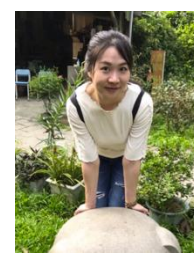

**Zih-Ling Chen** is currently working toward the master degree in the Graduate Institute of Brain and Mind Sciences, National Taiwan University College of Medicine. She is an occupational therapist dedicated to providing customized intervention programs to improve a patient's ability to perform daily activities and their various life roles. Current research concerns sustained attention and mind wandering revealed by eye movement patterns and event-related potential.

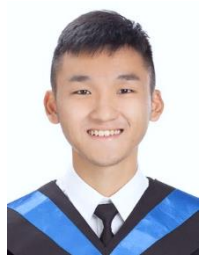

**Win-Ken Beh** received the B.S. degree in Electrical Engineering from National Taiwan University in 2018. He is currently working toward Ph.D. degree from the Graduate Institute of Electronic Engineering, National Taiwan University, Taipei, Taiwan. His research interests are in the areas of compressive sensing, DSP algorithm for bio-signal and signal quality evaluation for wearable applications.

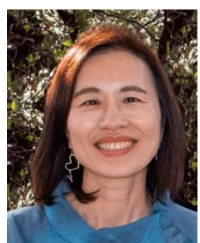

**Su-Ling Yeh** received her B.S. and M.S. degrees in Psychology from National Taiwan University (NTU), Taiwan, and Ph.D. degree in cognitive psychology from the University of California, Berkeley, USA. Since 1994, she has been with the Department of Psychology, NTU and was awarded Lifetime Distinguished Professorship in 2012. She is a recipient of Academic award of Ministry of Education and Distinguished Research

Award of National Science Council of Taiwan. She is an APS (American Psychological Science) fellow, and 2019-20 Stanford-Taiwan Social Science Fellow at the Center for Advanced Study in the Behavioral Sciences, Stanford University. She serves as associate director of NTU Center for Artificial Intelligence and Advanced Robotics and Editorial Board of Scientific Reports. Her research interests include cognitive neuroscience, perception, attention, consciousness, multisensory integration, aging, and applied research on display technology, eye tracking device, affective computing, and AI/robots.

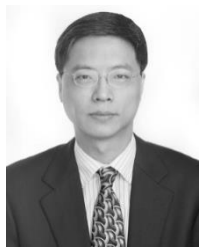

**An-Yeu (Andy) Wu (M'96-SM'12-F'15)** received the B.S. degree from National Taiwan University in 1987, and the M.S. and Ph.D. degrees from the University of Maryland, College Park in 1992 and 1995, respectively, all in Electrical Engineering. In August 2000, he joined the faculty of the Department of Electrical Engineering and the Graduate Institute of Electronics Engineering, National Taiwan University, where he is currently a distinguished professor. His research interests include VLSI architectures for signal processing and communications, and adaptive/multirate signal processing. He has published more than 190 refereed journal and conference papers in above research areas, together with five book chapters and 16 granted US patents. From August 2007 to Dec. 2009, he was on leave from NTU and served as the Deputy General Director of SoC Technology Center (STC), Industrial Technology Research Institute (ITRI), Hsinchu, Taiwan. In 2010, he received "Outstanding EE Professor Award" from The Chinese Institute of Electrical Engineering (CIEE), Taiwan. From 2012 to 2014, he served as the Chair of VLSI Systems and Applications (VSA) Technical Committee (TC), one of the largest TCs in IEEE Circuits and Systems (CAS) Society. In 2015, he is elevated to IEEE Fellow for his contributions to DSP algorithms and VLSI designs for communication IC/SoC.